\documentclass[%
 reprint,
 amsmath,amssymb,
 aps,
]{revtex4-2}

\usepackage{graphicx}
\usepackage{dcolumn}
\usepackage{bm}

\usepackage{xcolor}

\begin{document}


\title{Detecting stable adsorbates of (1S)-camphor on Cu(111) with Bayesian optimization}

\author{Jari J\"{a}rvi}
\email[]{jari.jarvi@aalto.fi}
\author{Patrick Rinke}
\author{Milica Todorovi\'{c}}
\affiliation{Department of Applied Physics, Aalto University, P.O. Box 11100, 00076 Aalto, Espoo, Finland}

\begin{abstract}
Identifying the atomic structure of organic--inorganic interfaces is challenging with our current research tools. Interpreting the structure of complex molecular adsorbates from microscopy images can be difficult, and using atomistic simulations to find the most stable structures is limited to partial exploration of the potential energy surface due to the high-dimensional phase space. In this study, we present the recently developed Bayesian Optimization Structure Search (BOSS) method as an efficient solution for identifying the structure of non-planar adsorbates. We apply BOSS with density-functional theory simulations to detect the stable adsorbate structures of (1S)-camphor on the Cu(111) surface. We identify the optimal structure among 8 unique types of stable adsorbates, in which camphor chemisorbs via oxygen (global minimum) or physisorbs via hydrocarbons to the Cu(111) surface. This study demonstrates that new cross-disciplinary tools, like BOSS, facilitate the description of complex surface structures and their properties, and ultimately allow us to tune the functionality of advanced materials.
\end{abstract}

\maketitle


\section{Introduction}

Our frontier technologies are increasingly based on advanced functional materials, which are often blends of organic and inorganic components. For example, in search for renewable energy solutions, hybrid perovskites are currently the best candidate to replace silicon in our solar cells \cite{Snaith2018}. In medicine, hybrid materials have been studied extensively for applications in tissue engineering \cite{Terzaki2013} and drug delivery \cite{Contessotto2009}. To optimize the functional properties of these materials, we need detailed knowledge of their atomic structure. Of particular interest is the hybrid interface, which has a central role in defining the electronic properties of the material. 

Assemblies of organic molecules on surfaces have been studied experimentally, for example with x-ray diffraction \cite{Farronato2018, Wang2020}, scanning tunneling microscopy \cite{Kuhnle2009, Kumar2017, Guo2017} and atomic force microscopy (AFM) \cite{Wold2002, Monig2018, Zint2017}. These methods have a considerable resolution in imaging planar surface structures, but interpreting images of bulky 3-dimensional molecules on surfaces can be difficult, which prevents an accurate structure determination. In such cases, computations can help in detecting the most stable structures.

With atomistic simulations, we can determine the optimal structures by computing the potential energy surface (PES). We can identify the stable structures in the minima of the PES and evaluate their mobility via the associated energy barriers. The most stable structure, that is the most probable structure in nature, corresponds to the global minimum of the PES. For its reliable identification, we must explore the PES thoroughly.

Calculating the full PES with complex hybrid materials requires either i) fast energy computations, or ii) an advanced method of constructing the complete PES with a small number of energy points. Classical force-field potentials are fast to compute, but they cannot accurately model hybrid materials, in which atomic interactions often feature a mixture of covalent and dispersive bonding, with charge transfer and polarization effects. Instead, we must employ quantum mechanical methods, such as density-functional theory (DFT) \cite{Hohenberg1964, Kohn1965}, for electronically accurate energy sampling. With hybrid materials, this makes thorough exploration of the PES prohibitively expensive with conventional phase-space exploration methods, such as minima hopping \cite{Goedecker2004}, Monte Carlo methods \cite{Li2012}, or metadynamics \cite{Laio2002}, which typically require calculating thousands of energy points on the PES.

Traditionally, the stable structures have been identified by initializing the minima search with estimated low energy structures, based on chemical intuition \cite{Obersteiner2017, Packwood2017a}, thus narrowing down the search space. With hybrid materials, however, this intuition is difficult to apply and can lead to biased or incorrect results.  For example, with only partial knowledge of the PES, a metastable local minimum energy structure could easily be misinterpreted as the most stable global minimum.

Recently, Gaussian processes (GPs) \cite{Rasmussen2006} and Bayesian optimization (BO) \cite{Shahriari2016} have been applied in modeling the PES to identify minimum-energy structures. GP regression has been used for example in local structure optimization \cite{GarijodelRio2019}, in finding minimum energy paths \cite{Koistinen2017}, and in predicting specific materials properties, such as melting temperature \cite{Seko2014} or elasticity \cite{Balachandran2016}. BO has been applied in detecting molecular conformers \cite{Chan2019} and adsorbate structures \cite{Packwood2017, Carr2016}, in identifying stable molecular compounds \cite{Jorgensen2018}, and in discovering materials with low thermal hysteresis \cite{Xue2016} or thermal conductivity \cite{Seko2015}. Typically, previous studies have employed customized material-specific models, using for example a coarse-grained search space with discrete molecular configurations, or predetermined GP hyperparameters, at the cost of generality of the method.

In this work, we show that the recently developed Bayesian Optimization Structure Search (BOSS) machine-learning method \cite{Todorovic2019, Egger2020, Fang2020, BOSSwebsite} provides a solution to the structure search conundrum. With BOSS, we adopt the aforementioned approach ii) and construct the complete PES using a small number of energy points. To demonstrate the capabilities of BOSS, we apply it with DFT to (1S)-camphor (C$_{10}$H$_{16}$O, hereafter shortened as camphor) adsorption on the Cu(111) surface. Camphor is an exemplary case of a bulky molecule, which is difficult to image with microscopy. AFM experiments \cite{Alldritt2020} have revealed various different conformers of camphor on Cu(111), which makes it ideal for benchmarking the BOSS method.  

Our objective is to detect the stable adsorbate structures of camphor on Cu(111). With BOSS, we build a surrogate model of the PES of adsorption and data-mine this PES to identify the stable structures in its minima. We converge the model for a reliable detection of all the PES minima, not only the global energy minimum. We estimate the mobility of the adsorbates from the energy barriers extracted from the surrogate PES and analyze the electronic structure of each adsorbate. Our results provide insight into the adsorption of complex organic molecules on metallic substrates and pave the way to more complex studies of hybrid monolayer formation and hybrid interfaces.

In the following sections, we first introduce our computational methods for adsorbate structure identification with BOSS, the first-principles calculations, and their application on detecting the stable adsorbates of camphor on Cu(111). We then present our results, discuss our findings, and conclude the analysis.

\section{Computational methods}
\label{sec:compmeth}

\subsection{Adsorbate structure identification}

\begin{figure}
\includegraphics{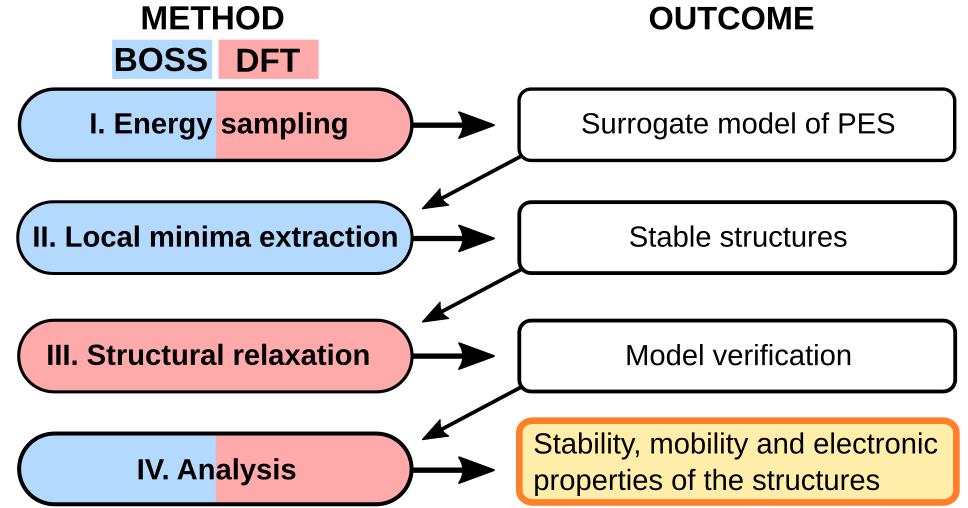}
\caption{\label{fig_process}Structure search with BOSS (blue) and DFT (red). I) The PES is sampled with BOSS by calculating energies of atomic configurations with DFT to obtain the surrogate model of the PES. II) BOSS identifies the stable structures in the minima of the PES. III) The stable structures are confirmed with full relaxation with DFT, after which IV) their mobility and adsorption properties are analyzed via the corresponding energy barriers and electronic structure.}
\end{figure}

BOSS is a machine-learning method that accelerates structure search via strategic sampling of the PES. With given initial data, BOSS builds the most probable surrogate model of the PES, refines it iteratively with active learning, and identifies the stable structures in the minima or the PES. In this work, we apply BOSS with DFT for accurate sampling of the energy points. In the following, we introduce the 4-step process (Fig.~\ref{fig_process}) of structure detection with BOSS and DFT, in analogy to Ref.~\cite{Todorovic2019}. We construct the surrogate model of the PES by sampling the adsorption energies with DFT (I). We then identify the stable structures by extracting the local minima of the PES (II) and verify them with full structural relaxation with DFT (III). We analyze the relaxed structures (IV) for their stability and mobility via the energy barriers on the PES, and investigate their electronic properties with DFT.

\subsubsection{Bayesian Optimization Structure Search}

With the atomic structures and their corresponding energies, BOSS constructs a surrogate model of the PES. We define the atomic structures using chemical \emph{building blocks} \cite{Oganov2006}, which are natural rigid components of the structure, for example rigid molecules, functional groups, or a surface slab. The PES is then defined in the phase space resulting from the remaing degrees of freedom, for example the relative translation and/or rotation of building blocks.

\begin{figure*}
\includegraphics{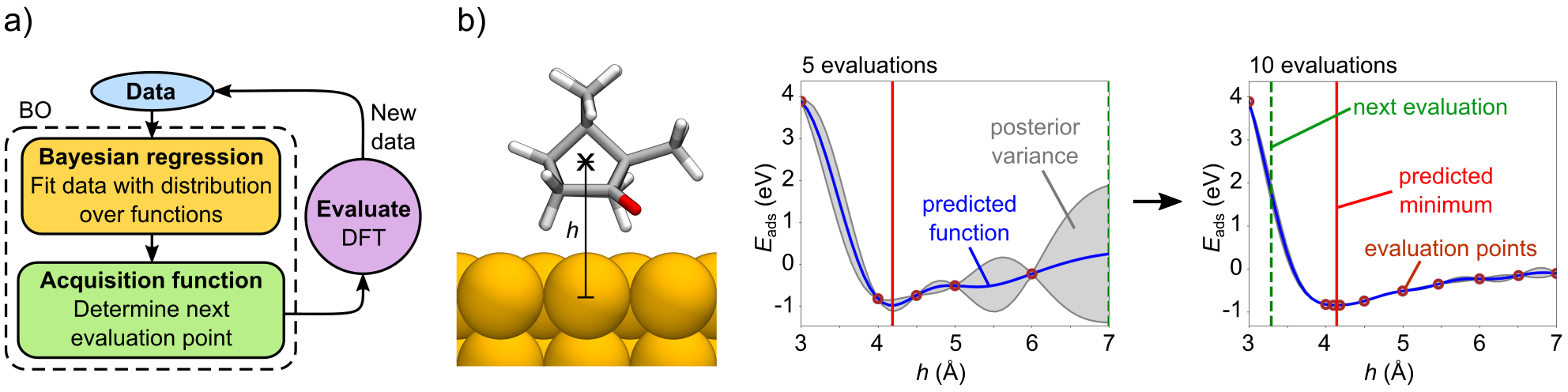}
\caption{\label{fig_boss}BOSS workflow and example performance. a) Basic principle of the BOSS method, in which Bayesian optimization (BO) is applied iteratively with DFT to build a surrogate model of the PES. b) 1D example of the iterative process, in which the adsorption energy $E_{\mathrm{ads}}$ of camphor on Cu(111) is predicted as a function of height $h$ of the molecule from the surface. The predicted adsorption height is converged in 5 energy evaluations to within 0.1~\AA. After 10 evaluations, the posterior variance, which describes the uncertainty of the model, has become vanishingly small throughout the search region.}
\end{figure*}

BOSS refines the PES model iteratively with active learning using BO (Fig.~\ref{fig_boss}a). We here only sketch the search principle and refer the interested reader to a more in-depth presentation and to the theoretical foundation in Refs.~\cite{Rasmussen2006,Gutmann2016,Todorovic2019}.  BO is a two-step process, in which data is first fitted with a GP distribution over functions using Bayesian regression. With the resulting surrogate model (Fig.~\ref{fig_boss}b), BOSS then determines the next sampling point using an acquisition function.

In the surrogate model, the posterior mean is the most probable model of the predicted function (here the PES). The posterior variance describes the uncertainty of the model in less explored areas. It therefore vanishes at the known data points. 

The next sampling point is determined using the exploratory Lower Confidence Bound (eLCB) \cite{Brochu2010} acquisition function, which balances \emph{exploitation} against \emph{exploration}. In exploitation, BO refines the model by acquiring the next point near the currently predicted global minimum. In exploration, the next acquisition is made at the point of maximum posterior variance, exploring less visited areas. In this study, we converge the PES model with respect to the coordinates and energy of all the minima, not only the global energy minimum.

\subsubsection{Local minima and barrier extraction}

Once the PES is converged, we data-mine the surrogate model. We extract the lowest energy minima, which we equate with the lowest energy adsorbate structures. The minima are detected using the built-in local minima search functionality of BOSS. The search is performed with minimizers, which apply the limited-memory Broyden–Fletcher–Goldfarb–Shanno (L-BFGS) \cite{Byrd1995} optimization algorithm. The minimizers start in different regions of the PES and traverse the landscape, following the gradients to locate the minima. 

The confidence of the surrogate model in different regions of the PES is quantified via the standard deviation ($\sigma^\textrm{B}$), which is the square root of the posterior variance in the GP model (Fig.~\ref{fig_boss}b). With the standard deviation, we evaluate the confidence of the surrogate model in the identified minima. Furthermore, we evaluate the accuracy of the model by computing the energy of each identified minima structure with DFT ($E^\textrm{D}$) and compare it to the corresponding energy in the surrogate model ($E^\textrm{B}$).

The BOSS PES also provides access to energy barriers, with which we can estimate the mobilities of our identified adsorbate structures. BOSS provides post-processing tools to locate the lowest energy barriers between two minima in the PES with the nudged elastic band (NEB) method. However, since we compute the PES with the building block approximation and not in the space of all atomic degrees of freedom, these energy barriers are only upper limits to the true barriers. However, even qualitative accuracy in barrier evaluations suffices to identify the least mobile structures, which are the best candidates when compared to experimentally observed structures. We will return to energy barriers and our way of estimating them in the results section, after we have introduced the camphor/Cu(111) system in more detail.

\subsubsection{Structural relaxation and analysis}

We verify the identified structures against a full DFT structure relaxation. In this, we remove the building block approximation and allow unrestricted motion of all atoms according to the inter-atomic forces in DFT. We then quantify and analyze the structural changes in the relaxation with respect to the atomic coordinates and the energy change ($\Delta E^\textrm{D}_\textrm{R}$) for each structure.

To validate the building block approximation, we evaluate the changes in the internal geometry of the building blocks after releasing them in the relaxation. For this, we calculate the average root-mean-square deviation of the atomic positions and the mean deviation of bond lengths, comparing the structures before and after the relaxation. 

We furthermore investigate the electronic structure of the stable adsorbates by analyzing their partial density of states (DOS) and the charge distribution with the Mulliken analysis of partial charges \cite{Mulliken1955}.

\subsection{Camphor on Cu(111)}

\begin{figure*}
\includegraphics{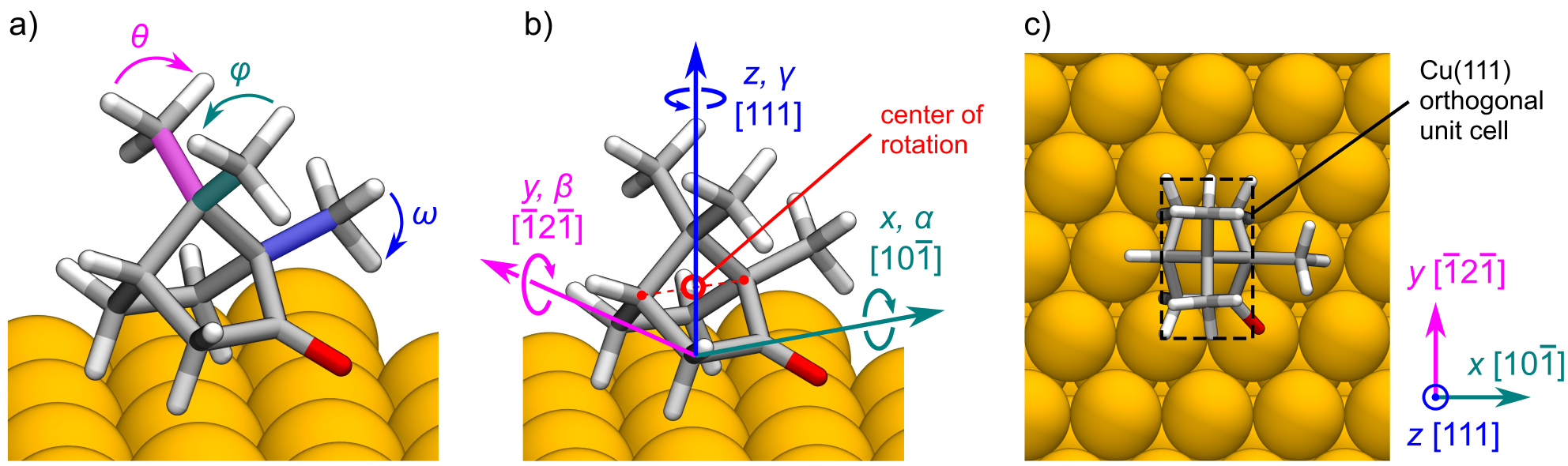}
\caption{\label{fig_dof}Degrees of freedom for the minimum energy search. a) Three methyl group rotation angles $\theta$, $\varphi$ and $\omega$ of camphor in the 3D conformer search with BOSS. b) Three translational directions ($x$, $y$, $z$) and three rotation angles ($\alpha$, $\beta$, $\gamma$) of camphor in the 6D search for stable adsorbate structures. The center of rotation is the middle point between the two carbon atoms, highlighted in red. c) Orthogonal unit cell of Cu(111), which is the search range in $x$ and $y$ directions.}
\end{figure*}

We study the adsorption of camphor on the Cu(111) surface using 2 building blocks: i) the global minimum camphor conformer and ii) the Cu(111) surface slab. With BOSS, we first identify the global minimum camphor conformer without the Cu(111) surface with a 3D search of methyl group rotations (Fig.~\ref{fig_dof}a). We normalize the lengths and angles of the C-H bonds in the 3 methyl groups by setting them to identical values, based on their average lengths and angles (see supplementary material (SM) Fig.~\ref{fig_sm_ideal}). With this, we obtain an ideal camphor geometry with 3 identical minima in the methyl group rotation (i.e. $120^\circ$ periodicity). We then study the rotation of the methyl groups in the ranges $\theta, \varphi, \omega \in [-60, 60]^\circ$.

With the identified global minimum conformer, we study the adsorption of camphor on Cu(111) with respect to molecular orientation and location. We define the PES of adsorption in a 6D phase space with 3 rotational angles ($\alpha, \beta, \gamma$) and 3 translational directions ($x, y, z$), which correspond to Cu lattice directions \mbox{[10-1]}, \mbox{[-12-1]} and \mbox{[111]}, respectively (Fig.~\ref{fig_dof}b). The adsorption height of the molecule ($z$) is defined with respect to the center point of rotation (Fig.~\ref{fig_dof}b), which is the middle point of the line connecting two C atoms at the sides of the rigid cage of camphor. We investigate the orientation of the molecule with full $360^\circ$ rotation of all angles, in the range $(\alpha, \beta, \gamma) \in [-180, 180]^\circ$. The search range on the $x$-$y$ plane of the Cu(111) surface is $(x,y) \in [-0.5, 0.5]$ (Fig.~\ref{fig_dof}c), defined in fractional unit-cell coordinates, which corresponds to lattice vectors $[a', b'] = [2.57, 4.45]$~\AA. 

Before we embarked on the full 6D camphor on Cu(111) search, we first scanned the system with several lower dimensional searches. Such low-dimensional searches (e.g. 1D variation of the adsorption height or 2D scans of molecular registry on the surface) permit us to relatively quickly explore the behavior of the system. We use them to find appropriate limits for the search dimensions (e.g. maximum and minimum height over the surface). Additionally, low-dimensional simulations help us to assess the contributions from rotational and translational degrees of freedom separately, to estimate the expected number of local minima and their approximate values, and to develop qualitative checks for expected energy landscapes (e.g. reflecting surface symmetries). The computational effort associated with these preparatory simulations is recycled, since all points sampled in reduced dimensions later serve as input in the 6D study. We note that analysis of low-dimensional simulations provides us with  qualitative insight into surface adsorpton, quantitative conclusions on the stable structures can only be drawn from a full 6D search.

\begin{figure*}
\includegraphics{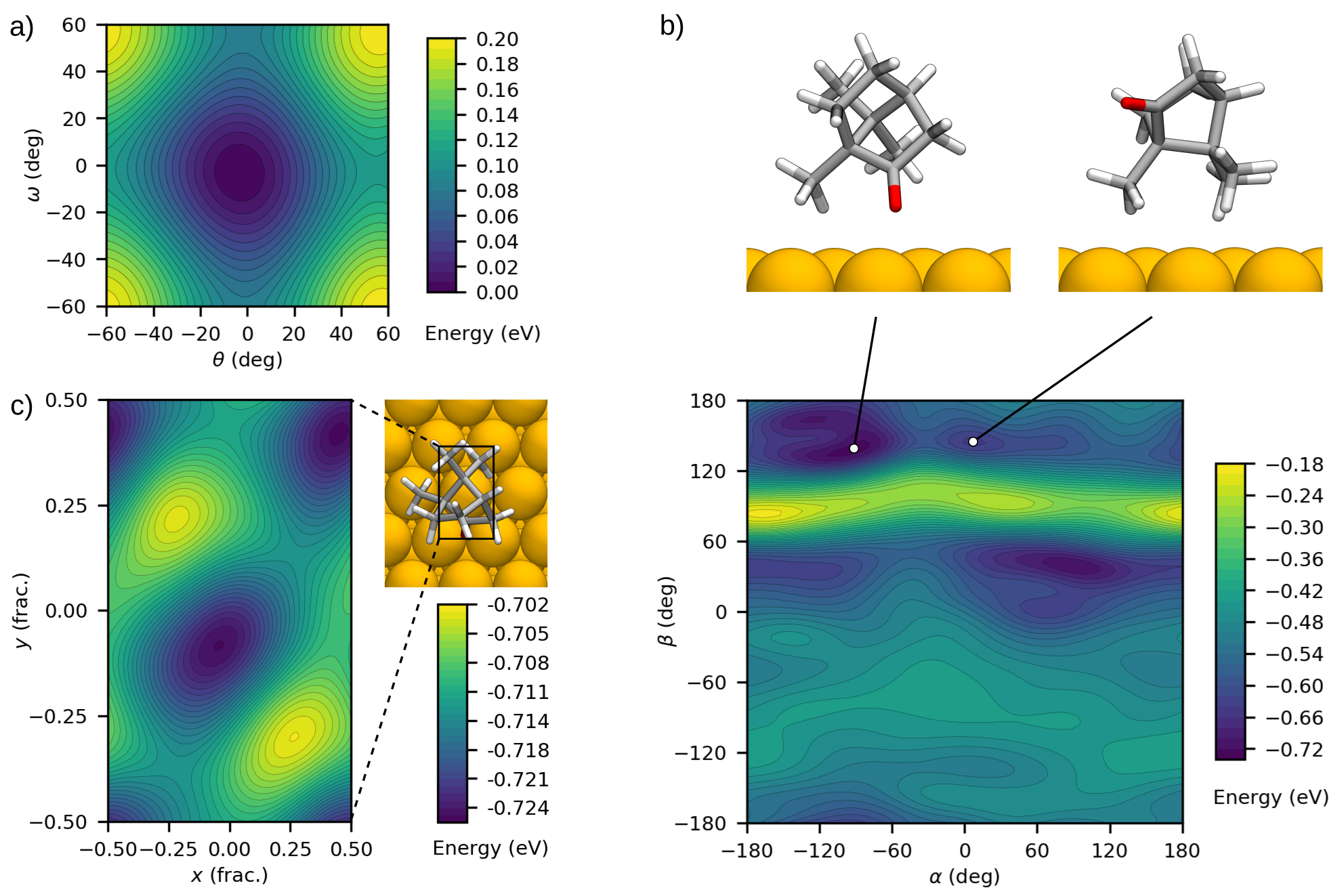}
\caption{\label{fig_land}Energy landscapes from preparatory BOSS simulations. a) $\theta$-$\omega$ 2D cross-section of the 3D PES in the camphor conformer search, featuring a single minimum and an energy barrier of 0.1 eV for methyl group rotation. b) $\alpha$-$\beta$ 2D cross-section of the 3D PES in the search for adsorption orientation of camphor on Cu(111). The landscape features multiple local minima and a higher-energy region at $\beta \approx 90^{\circ}$. c) PES of the 2D translational $x$-$y$ search of the adsorption site of camphor on Cu(111). The landscape has two identical minima, which agree with the translational symmetry in the orthogonal unit cell.}
\end{figure*}

With BOSS, we perform 3 low-dimensional searches, in which we study the adsorption of camphor on Cu(111) as a function of its i) adsorption height (1D), ii) orientation (3D), and iii) adsorption site (2D). First, we investigate the height of the molecule with a 1D search (Fig.~\ref{fig_boss}b) to determine a suitable height for the rotational search. Based on the resulting energy curve we estimate the optimal height  at which we avoid high energy peaks in all molecular orientations, and conduct the 3D rotational search. We then set the molecule in the observed minimum energy orientation (Fig.~\ref{fig_land}b) and perform a 2D search of the adsorption site within the orthogonal unit cell of the Cu(111) surface (Fig.~\ref{fig_dof}c). With the acquired knowledge of the energy ranges, we then determine the optimal height range of the molecule for the 6D search.

We perform a 6D search with combined degrees of freedom to identify the stable adsorbate structures of camphor on Cu(111). The search is initialized using the previously acquired energy points from the low-dimensional studies. We multiply the number of initial energy points by applying the 2-fold translational symmetry in the orthogonal unit cell and the 3-fold rotational symmetry of the Cu(111) surface at the on-top site. With BOSS, we acquire new energy points and converge the 6D PES with respect to the energy and coordinates of the identified local minima (details provided in SM).

The electronic structure of the stable adsorbates is analyzed with the partial DOS and the Mulliken analysis of partial charges. We compare the partial DOS of the adsorbed camphor to the highest occupied and lowest unoccupied molecular orbitals (HOMO and LUMO, respectively) of an isolated camphor molecule. In the Mulliken analysis, we calculate the sum of partial atomic charges per element in the adsorbed camphor and compare them to the corresponding charge distribution of an isolated molecule. With this analysis, we study the effect of adsorption on the electronic structure of camphor in the identified stable structures.

\subsection{First-principles calculations} 
\label{sec:DFT}

We use density-functional theory to calculate the adsorption energy of camphor on Cu(111) in the BOSS runs, to relax the predicted stable structures and to analyze the electronic structure of the stable adsorbates. We apply the all-electron, numeric atom-centered orbital code FHI-aims \cite{Blum2009,Havu2009,Ren2012} with the Perdew-Burke-Ernzerhof (PBE) exchange-correlation functional \cite{Perdew1996}. PBE is augmented with van der Waals (vdW) corrections employing the vdW$^\mathrm{surf}$ parametrization \cite{Ruiz2016} of the Tkatchenko-Scheffler method \cite{Tkatchenko2009}. Previous work found that PBE+vdW$^\mathrm{surf}$ adequately describes organic molecules on metal surfaces  \cite{Ruiz2016,Hofmann2013,Ruiz2012}.

Our converged settings employ \emph{Tier 1} basis sets with \emph{light} grid settings and a $\Gamma$-centered $3 \times 2 \times 1$ k-point mesh with a $(6 \times 4) \sqrt{3}$ supercell model. We apply relativistic corrections with the zero-order regular approximation \cite{vanLenthe1993} and Gaussian broadening of 0.1~eV of the electronic states. The total energy is converged within $10^{-6}$~eV in the self-consistency cycle and the structures are relaxed below a maximum force component of $10^{-2}$~eV/\AA. 

We model the Cu substrate with a Cu(111) slab of 4 atomic layers and $(6 \times 4) \sqrt{3}$ orthogonal unit cells (192 atoms, lattice vectors $[a, b, c] = [15.41, 17.79, 56.29]$~\AA). The lattice constant of Cu is set to 3.632~\AA, which we obtain from relaxed bulk Cu, in agreement with reference studies \cite{Haas2009, Liu2013}. We construct the 4-layer Cu slab by fixing the two bottom layers to their optimal layer separation ($d_{34} = 2.097$~\AA, corresponding to bulk Cu). The two top layers are then relaxed, which results in a reduced layer separation ($d_{12} = 2.076$~\AA, $d_{23} = 2.081$~\AA), in agreement with previous calculations \cite{Stroppa2007}. We apply this Cu slab model as a building block in the subsequent study of camphor adsorption. 

Our other building block is the global minimum conformer of camphor, which we add onto the Cu slab model. The $(6 \times 4) \sqrt{3}$ supercell provides a good approximation of a single molecule on the surface, with average lateral separation of 10~\AA~between the periodic images of camphor and 50~\AA~separation between the periodic Cu(111) slabs. 

The adsorption energy $E_\mathrm{ads}$ is calculated as

\begin{equation}
E_\mathrm{ads} = E_\mathrm{tot} - (E_\mathrm{Cu} + E_\mathrm{cam}),
\end{equation}

\noindent in which $E_\mathrm{tot}$ is the total energy of the camphor/Cu(111) system, $E_\mathrm{Cu}$ the energy of the relaxed Cu slab and $E_\mathrm{cam}$ the energy of an isolated camphor molecule.

\section{Results}
\label{sec:results}

\subsection{Camphor conformer search}

We analyzed the camphor conformers with a 3D BOSS search of the methyl group rotations. The energy landscape (Fig.~\ref{fig_land}a), converged in 20 evaluations, features a single global energy minimum at $(\theta, \varphi, \omega) = (-3, 7, -3)^\circ$, and an energy barrier of 0.1~eV for the rotation of the methyl groups. Given this barrier, the rotation of the methyl groups $\Delta \varphi$ away from the global minimum is expected to be small at room temperature. The Arrhenius law predicts that in 50~\% of the molecules $\Delta \varphi < 10^\circ$, and in 70~\% $\Delta \varphi < 15^\circ$. Camphor is thus likely to be found in a conformation very close to the global minimum geometry. We thus take the identified conformer as a building block in the following adsorption study. Any further structural deformations are accounted for at a later stage with full DFT relaxation.

\subsection{Qualitative insight into adsorption of camphor on Cu(111)}

Before conducting a full 6D search, we carried out several lower-dimensional searches to develop a feeling for the behavior of camphor on Cu(111). First, we performed a 1D search in the $z$ direction, then a 3D rotational search in ($\alpha, \beta, \gamma$), and finally a 2D translational search on the $x$-$y$ plane.

We learned about the adsorption height range of camphor on Cu(111) from a 1D BOSS search (Fig.~\ref{fig_boss}b) within the limits $z \in [3, 7]$~\AA\ (other variables were set to $(x, y, \alpha, \beta, \gamma) = (0, 0, 0, 0, 0)$). The predicted minimum of the adsorption energy converged in 5 evaluations and is found at $-0.847$~eV at height $z = 4.14$~\AA. The energy curve has a strong dispersive character and the repulsive energy increases rapidly as the molecule approaches the surface below 4~\AA.

For the 3D rotational study, we placed the molecule into a fixed position at $(x, y, z)$ = (0, 0, 5)~\AA~to avoid close contact between the molecule and the surface. Molecular placement at the on-top site (above Cu atom) here allows us to curtail the $\gamma$ range to $[-60, 60]^\circ$. The resulting PES (Fig.~\ref{fig_land}b) converged in 115 evaluations and contains many features associated with different reactive sites of camphor. The higher energy band at $\beta \approx 90^\circ$ corresponds to the closest approach of the molecule to the surface (via methyl group $\omega$ in Fig.~\ref{fig_dof}a). The multiple minima and strong barriers imply that camphor may adsorb on Cu(111) in various stable orientations. We explored the structures associated with the most favourable minima to infer the binding mechanisms. As shown in Fig.~\ref{fig_dof}b, we found that both charge-withdrawing O and the neutral methyl groups face the surface, suggesting that both chemical and dispersive bonding can be expected in the full 6D search.

The 2D search in the $x$-$y$ plane allowed us to compute the translational energy landscape for camphor. We set the molecule to the global minimum orientation $(\alpha, \beta, \gamma) = (-84, 143, 3)^\circ$ from the previous rotational search, at $z = 5$~\AA. The PES (Fig.~\ref{fig_land}c) converged in 20 evaluations and features two identical minima at $(x, y) = (-0.05, -0.08)$ and $(0.45, 0.42)$ in fractional coordinates of the unit cell. These correspond to the translational symmetry of the Cu(111) surface in the orthogonal unit cell. We conclude that our model fitting is qualitatively correct even when the landscapes are very flat, as with this choice of parameters. The flat energy landscapes indicate that rotational degrees of freedom may influence adsorption more than translational ones, but this is best verified in 6D.

Based on the low-dimensional studies, we expect to find multiple stable adsorbate structures in the 6D search, with varying molecular orientations and both chemical and dispersive bonding. Given the observed energy ranges, we conclude that the optimal search range for the height of the molecule in the 6D search is $z \in [4, 7]$~\AA. The range is sufficiently broad to include all the minima and avoids high energy peaks in the closest approach of the molecule to the surface.

\subsection{Predicted stable adsorbates}

For the 6D search of stable adsorbates, we employed the 492 previously acquired energy points from the low-dimensional studies. These points were then multiplied according to the translational and rotational symmetries of the Cu(111) surface, which resulted in 986 initial energy points for the 6D search. We converged the 6D PES (details provided in SM) by acquiring 197 new points, for which we also applied the symmetries. The surrogate model of the 6D PES was then constructed with 1380 energy points.

\begin{figure}
\includegraphics{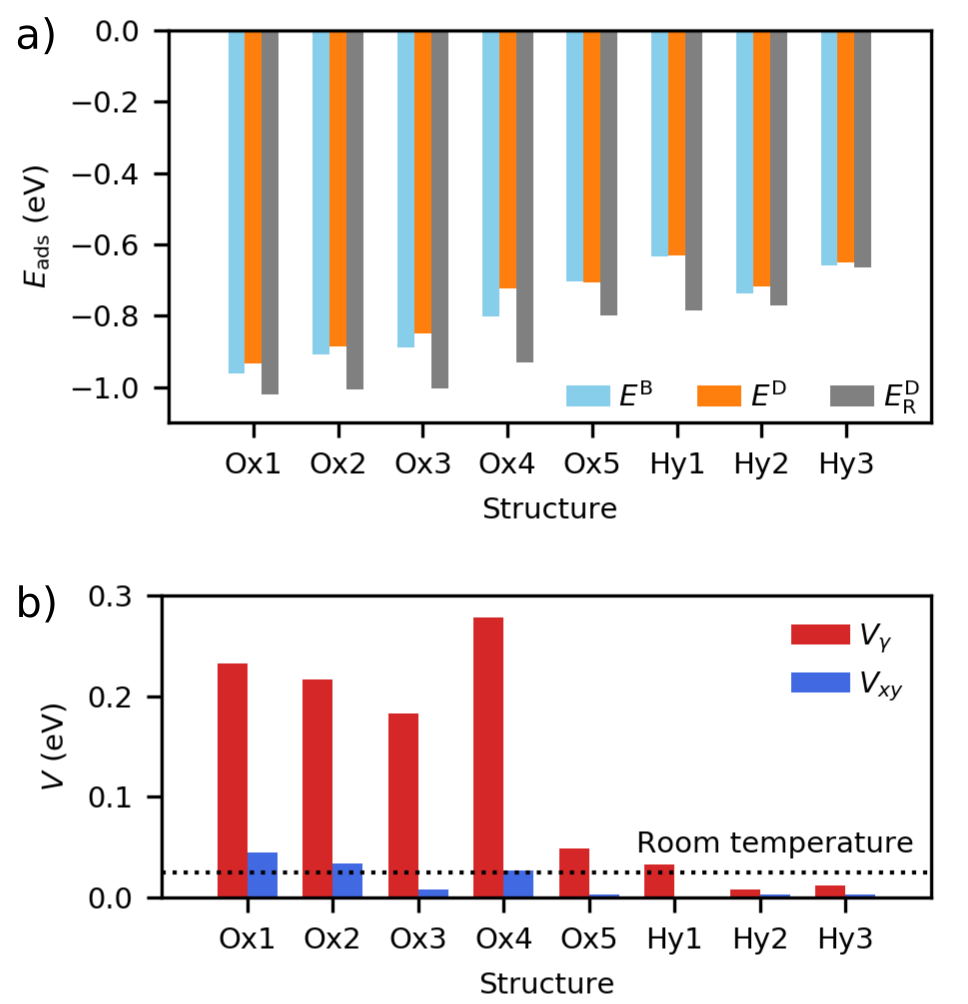}
\caption{\label{fig_energy}Energetics of adsorption and mobility for surface adsorbates. a) Adsorption energies ($E_\textrm{ads}$) of the stable adsorbates predicted by BOSS ($E^\textrm{B}$), their true values calculated with DFT ($E^\textrm{D}$) and the adsorption energies of the relaxed structures ($E^\textrm{D}_\textrm{R}$). b) Energy barriers ($V$) for $\gamma$ rotation ($V_\gamma$) and $x-y$ translation ($V_{xy}$), in comparison with thermal energy at room temperature.}
\end{figure}

In the minima of the PES, we identified 8 unique stable structures with predicted adsorption energies ($E^\textrm{B}$) in the range $[-0.961, -0.634]$~eV (Fig.~\ref{fig_energy}a and Tab.~\ref{tab_energies}). We have classified the structures with respect to the bonding species closest to the surface in the adsorbed camphor, namely oxygen (class Ox) and hydrogen (class Hy). The standard deviation of the adsorption energy ($\sigma^\textrm{B}$) in the surrogate PES is 0.019~eV in the global minimum and $0.025$~eV on average over all the minima (Tab.~\ref{tab_energies}), which shows low uncertainty of the model in these points. The energies of the identified structures, calculated with DFT ($E^\textrm{D}$) are in the range $[-0.933, -0.631]$~eV, in close agreement with the predicted energies.

\begin{table*}
\caption{\label{tab_energies}Adsorption energies of the stable adsorbates, predicted by BOSS ($E^\textrm{B}$), and their standard deviation in the surrogate model of the 6D PES ($\sigma^\textrm{B}$). Adsorption energies calculated with DFT ($E^\textrm{D}$) and their difference from the predicted energies ($\Delta E^\textrm{D}$). Energy after relaxation ($E^\textrm{D}_\textrm{R}$), and energy change in the relaxation ($\Delta E^\textrm{D}_\textrm{R}$). Predicted energy barriers of $\gamma$ rotation ($V_\gamma$) and $x$-$y$ translation ($V_{xy}$).}
\setlength{\tabcolsep}{7.5pt}
\begin{tabular}{ccc|cccc|cc}
 \hline
& $E^\textrm{B}$~(eV) & $\sigma^\textrm{B}$~(eV) & $E^\textrm{D}$~(eV) & $\Delta E^\textrm{D}$~(eV) & $E^\textrm{D}_\textrm{R}$~(eV) & $\Delta E^\textrm{D}_\textrm{R}$~(eV) & $V_\gamma$~(eV) & $V_{xy}$~(eV) \\
 \hline
Ox1 & $-0.961$ & $0.019$ & $-0.933$ & $+0.028$ & $-1.022$ & $-0.089$ & 0.232 & 0.045 \\
Ox2 & $-0.910$ & $0.013$ & $-0.885$ & $+0.025$ & $-1.008$ & $-0.123$ & 0.216 & 0.034 \\
Ox3 & $-0.889$ & $0.027$ &  $-0.850$ & $+0.039$ & $-1.005$ & $-0.155$ & 0.183 & 0.008 \\
Ox4 & $-0.803$ & $0.032$ &  $-0.723$ & $+0.079$ & $-0.932$ & $-0.209$ & 0.278 & 0.027 \\
Ox5 & $-0.704$ & $0.016$ &  $-0.706$ & $-0.002$ & $-0.800$ & $-0.094$ & 0.048 & 0.003 \\
\hline
Hy1 & $-0.634$ & $0.021$ &  $-0.631$ & $+0.003$ & $-0.784$ & $-0.154$ & 0.033 & 0.001 \\
Hy2 & $-0.737$ & $0.041$ &  $-0.719$ & $+0.019$ & $-0.772$ & $-0.053$ & 0.008 & 0.003 \\
Hy3 & $-0.658$ & $0.027$ &  $-0.652$ & $+0.005$ & $-0.664$ & $-0.012$ & 0.012 & 0.003 \\
\hline
\end{tabular}
\end{table*}

\subsection{Relaxed structures}

\begin{figure*}
\includegraphics{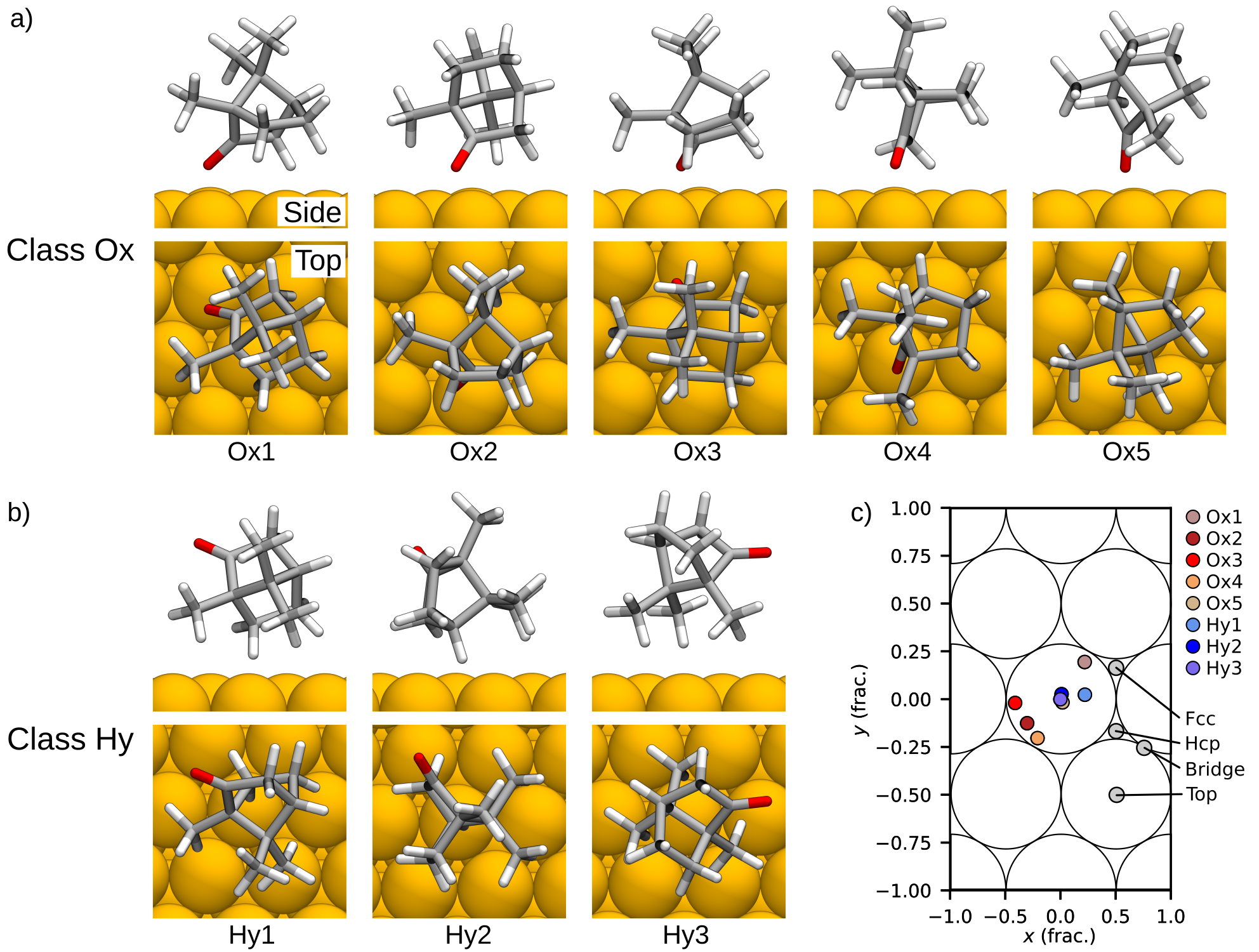}
\caption{\label{fig_structures}Relaxed stable adsorbate structures of camphor on Cu(111) in the 6D search, showing chemisorption of the molecule via oxygen (a), class Ox) and physisorption via hydrocarbons (b), class Hy). c) Adsorption site of camphor in the relaxed structures (center of the molecule) and the high symmetry points of the Cu(111) surface.}
\end{figure*}

We verified the stable structures by performing full DFT relaxations (Fig.~\ref{fig_structures}a and b). In the relaxation, we observed an average decrease of $-0.11$~eV from the $E^\textrm{D}$ energies (Fig.~\ref{fig_energy}a and Tab.~\ref{tab_energies}). We found that in class Ox structures, 80 \% of the binding energy is due to dispersion whereas in class Hy, the binding energy is purely dispersive.

The structural changes in the relaxation were analyzed by comparing the location and orientation of the molecule before and after the relaxation. We observed the relaxed structures to be almost identical with the initial ones. The average change in the location of the molecule, over all structures, is $(\overline{\Delta x}, \overline{\Delta y}, \overline{\Delta z}) = (0.13, 0.09, 0.19)$~\AA~and in the orientation $(\overline{\Delta \alpha}, \overline{\Delta \beta}, \overline{\Delta \gamma}) = (6.1, 5.8, 2.5)^\circ$. The structural changes in the Cu slab are minimal. The changes in the internal geometry of camphor in the relaxation, after removing the building block approximation, were evaluated using the average root-mean-square deviation of the atomic positions and the mean deviation of bond lengths, which are 0.13~\AA~and 0.003~\AA, respectively, on average over all structures (see SM for structure-specific data).

We analyzed the adsorption site of camphor in the relaxed structures (Fig.~\ref{fig_structures}c) with respect to the center of the molecule (Fig.~\ref{fig_dof}b). The adsorption sites show a notable difference between the two classes. Class Hy structures (in particular Hy2 and Hy3) prefer to adsorb close to the on-top site, whereas class Ox structures feature more variance in their location. 3 of class Ox structures (Ox1, Ox3, and Ox4) adsorb near the bridge site and Ox5 is close to the on-top site.

To estimate camphor mobility on the surface, we inspect translational and rotational barriers. The translational energy barriers were computed using 2D $x$-$y$ cross-sections (grid of $100 \times 100$ points) of the predicted 6D PES, as described in the methods section. For the $\gamma$ rotation barriers, we extracted 1D $\gamma$ energy profiles from the 6D PES but found them overly smooth and free of features expected for an asymmetric molecule rotating on the Cu(111) surface. We concluded the upper limits for $\gamma$ rotation to be too inaccurate and analyze the $\gamma$ energy barriers using the fully relaxed structures of local minima geometries. For each mininimum type, we rotated camphor in-place (center point of rotation in Fig.~\ref{fig_dof}b) and computed the rotational energy profile with a 1D BOSS search (converged in 15 evaluations). While this approach is still approximate, the resulting energy profiles exhibit features that correctly reflect surface symmetry and provide us with a better estimate of the barriers without investing time and computational expense into NEB calculations.

The predicted energy barriers of $\gamma$ rotation and $x$-$y$ translation (Fig.~\ref{fig_energy}b and Tab.~\ref{tab_energies}) are in the range [0.008, 0.278] and [0.001, 0.045]~eV, respectively. The barriers are highest in class Ox structures, specifically in structures Ox1--Ox4, with a notable difference to class Hy. When we take into account the standard deviation of the adsorption energy in the surrogate model (Tab.~\ref{tab_energies}), the smallest energy barriers (of the order of 0.01 eV and below) are practically zero. This indicates free rotation of structures Hy2 and Hy3, and free diffusion of structures Ox3, Ox5, and Hy1--Hy3, even at low temperatures.

\subsection{Electronic structure}

\begin{figure}
\includegraphics{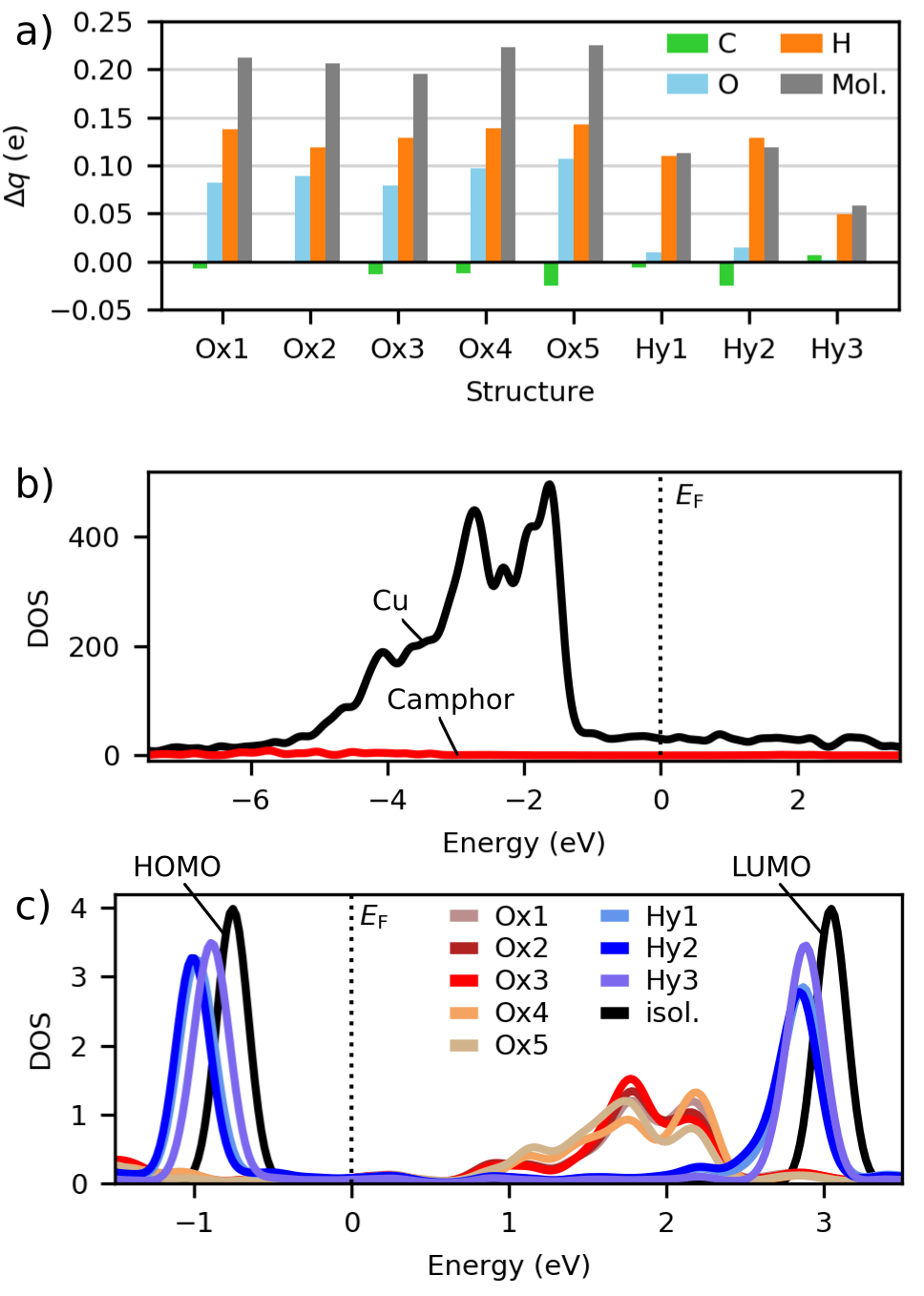}
\caption{\label{fig_charge}Electronic properties of different camphor adsorbates.
a) The sum of partial charges ($\Delta q$) in the adsorbed camphor in the relaxed structures. b) DOS of Cu and camphor in structure Ox1, and c) DOS of camphor in the relaxed structures and in an isolated molecule.}
\end{figure}

We analyzed the charge distribution of the stable adsorbates with the Mulliken analysis of partial charges and investigated their partial DOS to study the effect of adsorption on the electronic structure. The Mulliken analysis of partial charges in the relaxed structures (Fig.~\ref{fig_charge}a and Tab.~\ref{tab_charges}), in comparison to the charge distribution of an isolated camphor, shows electron transfer from the adsorbed camphor to the Cu substrate. The electron transfer is highest in class Ox structures, in which the O atom of camphor is close to the Cu surface. The average partial charge of camphor ($\Delta q$) is $+0.21$~e (elementary charge, $\textrm{e} = |\textrm{e}^-|$) in class Ox structures and $+0.10$~e in class Hy. In class Ox structures, the main contribution to the positive charge comes from hydrogen (H) atoms, with O as the second notable contributor. In class Hy structures, the positive charge of camphor originates predominantly from H atoms.

\begin{table}
\caption{\label{tab_charges}The sum of partial charges of C, O, and H in the adsorbed camphor ($\Delta q_\textrm{C}$, $\Delta q_\textrm{O}$, and $\Delta q_\textrm{H}$, respectively) and the total partial charge of camphor ($\Delta q$).}
\setlength{\tabcolsep}{7pt}
\begin{tabular}{ccccc}
 \hline
& $\Delta q_\textrm{C}$~(e) & $\Delta q_\textrm{O}$~(e) & $\Delta q_\textrm{H}$~(e) & $\Delta q$~(e) \\
 \hline
Ox1 & $-0.01$ & $+0.08$ & $+0.14$ & $+0.21$ \\
Ox2 & $-0.00$ & $+0.09$ & $+0.12$ & $+0.21$ \\
Ox3 & $-0.01$ & $+0.08$ & $+0.13$ & $+0.20$ \\
Ox4 & $-0.01$ & $+0.10$ & $+0.14$ & $+0.22$ \\
Ox5 & $-0.02$ & $+0.11$ & $+0.14$ & $+0.23$ \\
\hline
Hy1 & $-0.01$ & $+0.01$ & $+0.11$ & $+0.11$ \\
Hy2 & $-0.03$ & $+0.01$ & $+0.13$ & $+0.12$ \\
Hy3 & $+0.01$ & $+0.00$ & $+0.05$ & $+0.06$ \\
\hline
\end{tabular}
\end{table}

In the partial DOS of the relaxed structures (Fig.~\ref{fig_charge}b and c), we analyze the electronic states of the adsorbed camphor close to the Fermi level. The partial DOS of class Ox structures features hybridization of the electronic states, in comparison to the HOMO and LUMO of an isolated camphor. The hybridization implies chemical bonding between the molecule and the substrate in class Ox. Conversely, in class Hy, the electronic states resemble the HOMO and LUMO of an isolated camphor and are at $-1.0$ and $2.9$~eV, respectively, with an energy gap of $3.9$~eV. This indicates physisorption between the molecule and the substrate in class Hy.

\section{Discussion}
\label{sec:discussion}

With the low-dimensional studies of molecular translation (1D and 2D) and rotation (3D), we obtained a qualitative description of the adsorption properties of camphor on Cu(111). We gained insight on the estimated adsorption height of the molecule and acquired the ranges of adsorption energy with respect to molecular orientation and the adsorption site. The rotational energy landscape with multiple local minima suggests that camphor can adsorb on Cu(111) in various stable orientations. From the low-dimensional analysis, we obtained the required knowledge to determine the optimal search range for the height of the molecule in the subsequent 6D study.

In the relaxation of the identified structures, we observed minor changes in the molecular orientation, the adsorption site, and the adsorption energy. This effectively confirms the accuracy of the surrogate model of the 6D PES. Negligible changes in the internal structure of camphor and the Cu slab in the relaxation validates the building block approximation in this study. 

The 8 stable adsorbate structures extracted from the 6D search feature notable differences between the class Ox and Hy structures, specifically in their adsorption energy, adsorption site, energy barriers, and the electronic structure. Class Ox adsorbates have the highest adsorption energies and high energy barriers of molecular mobility. In class Ox, the preferred adsorption site is near the bridge site, so that O can point sideways to bond with the Cu atom. In class Hy, methyl groups avoid the on-top site, so the molecule centers there, and the methyl groups point sideways. The DOS of class Ox structures feature hybridization of the electronic states and the electron transfer from the molecule to the substrate is significantly larger than in class Hy, with the largest contribution per atom from O. This indicates chemisorption of camphor via O to the Cu substrate. Conversely, in class Hy structures we observed the characteristics of physisorption. Class Hy structures have systematically lower adsorption energies, energy barriers, and electron transfer to the substrate, and their DOS resemble the HOMO and LUMO of an isolated molecule. These findings are supported by the vdW contributions in the adsorption energy, which show 80~\% dispersive bonding in class Ox and fully dispersive adsorption in class Hy.

To verify the identified stable structures, we can compare them to adsorbates observed in experiments. The adsorption of camphor on Cu(111) has been studied experimentally with AFM by Alldritt et al. \cite{Alldritt2020}. In their images, they have observed various different adsorbate structures, which shows that camphor can adsorb on Cu(111) in multiple stable configurations. In the experiments, camphor molecules were deposited onto the Cu surface at 20~K temperature and the imaging was done at 5~K. When the surface is annealed to the imaging temperature, we expect the deposited molecules to obtain the global minimum conformer geometry, which corresponds to the camphor building block in this study. Based on the estimated energy barriers of molecular mobility in this study, we conclude that the experiments likely feature chemisorbed camphor molecules from class Ox. In particular, structures Ox1--Ox4, which have the highest barriers, are the most likely candidates for static adsorbates. They also have the highest adsorption energies, which makes them the most probable structures to be observed. Conversely, class Hy structures, which have lower adsorption energies and low energy barriers for molecular mobility, are less likely to be imaged in experiments. A more detailed comparison between BOSS and AFM will be reported in Ref.~\cite{Jarvi2020b}.

We highlight the computational efficiency of global structure search with BOSS by comparing the number of required DFT calculations to a conventional structure search. The best candidates for the minimum-energy structures can be first estimated using chemical intuition and then relaxed with DFT to identify the stable structures. With camphor on Cu(111), we can search for the stable adsorbates by placing the molecule on each of the 4 high symmetry points of the Cu surface (Fig.~\ref{fig_structures}c) and investigate for example 10 different molecular orientations at each of the adsorption sites. We estimate that the relaxation of the structures requires on average 40 calculation steps per structure. With this method, the estimated computational cost would be 1600 DFT calculations. Still, this amounts to exploring only a small portion of the PES and does not guarantee a reliable identification of the global minimum energy structure.

With BOSS, we identified the stable structures of camphor on Cu(111) with 892 DFT calculations (689 to construct the surrogate model of the 6D PES, and 203 to relax the 8 structures). Relaxation of the predicted stable structures in the local minima of the PES was fast (25 relaxation steps per structure on average) due to their low initial energy. With the PES model, we were able to reliably identify not only all the minima, but also the associated energy barriers of molecular mobility.  This comparison highlights the benefits of the BOSS approach, in particular i) computational efficiency, ii) reliable identification of the most stable structures, and iii) obtaining energy barriers readily with the surrogate model of the PES.

\section{Conclusion}
\label{sec:conclusion}

In this study, we have demonstrated the efficiency of BOSS in global structure search with complex molecular adsorbates. We have shown the accuracy of the constructed surrogate model of the PES, in comparison with adsorption energies of stable structures calculated with DFT. As a benchmark system, we have analyzed the adsorption of a camphor molecule on the Cu(111) surface with respect to molecular translation and rotation. With BOSS, we constructed a surrogate model of the 6D PES of adsorption and identified its minima, in which we detected the most stable structure (global minimum) and 7 other stable structures (local minima). 

We classified our stable structures into two classes Ox and Hy, with respect to the bonding species in the adsorbed camphor. The differences between the two classes were further categorized by the trends in the adsorption energies and the energy barriers of molecular motion. By analyzing the electronic structure of the stable adsorbates, we concluded that in the most stable structures (class Ox), camphor chemisorbs to the Cu surface via O bonding. Our results imply that class Ox structures are viable candidates for static camphor adsorbates observed in AFM experiments. 

By combining machine learning with DFT, BOSS provides a novel method for a reliable structure identification via the surrogate model of the PES. With the complete PES, we obtain chemical insight on numerous materials properties (e.g. the stable adsorbate structures and their mobility) in one go, without prior presumptions about the material. Our approach eliminates the human bias present in conventional structure search, in which the optimal structures are commonly estimated using chemical intuition. Efficient and unbiased structure search methods, such as BOSS, facilitate the study of complex hybrid interface structures. The acquired knowledge can be applied in precision-engineering interface structures in functional materials to optimize their advantageous properties.

\section*{acknowledgements}
The authors wish to acknowledge CSC -- IT Center for Science, Finland, and the Aalto Science-IT project for generous computational resources. An award of computer time was provided by the Innovative and Novel Computational Impact on Theory and Experiment (INCITE) program. This research used resources of the Argonne Leadership Computing Facility, which is a DOE Office of Science User Facility supported under Contract DE-AC02-06CH11357.

\section*{funding}
This work has received funding from the Academy of Finland via the Artificial Intelligence for Microscopic Structure Search (AIMSS) project No. 316601 the Flagship programme: Finnish Center for Artificial Intelligence FCAI, and from the Emil Aaltonen Foundation. 

\bibliography{references}


\clearpage
\onecolumngrid

\begin{center}
\textbf{\large Detecting stable adsorbates of (1S)-camphor on Cu(111) with Bayesian optimization \\
    \vspace{0.5em} SUPPLEMENTARY MATERIAL} \\
\vspace{1em}
Jari J\"{a}rvi,$^*$ Patrick Rinke, and Milica Todorovi\'{c} \\
\textit{Department of Applied Physics, Aalto University, P.O. Box 11100, 00076 Aalto, Espoo, Finland}
\end{center}

\setcounter{equation}{0}
\setcounter{section}{0}
\setcounter{figure}{0}
\setcounter{table}{0}
\setcounter{page}{1}
\makeatletter

\renewcommand{\theequation}{S\arabic{equation}}
\renewcommand{\thefigure}{S\arabic{figure}}
\renewcommand{\thetable}{S\Roman{table}}
\renewcommand{\bibnumfmt}[1]{[S#1]}
\renewcommand{\citenumfont}[1]{S#1}


\section{Camphor geometry in global minimum conformer search}

\begin{figure}[h]
\includegraphics[width=10cm]{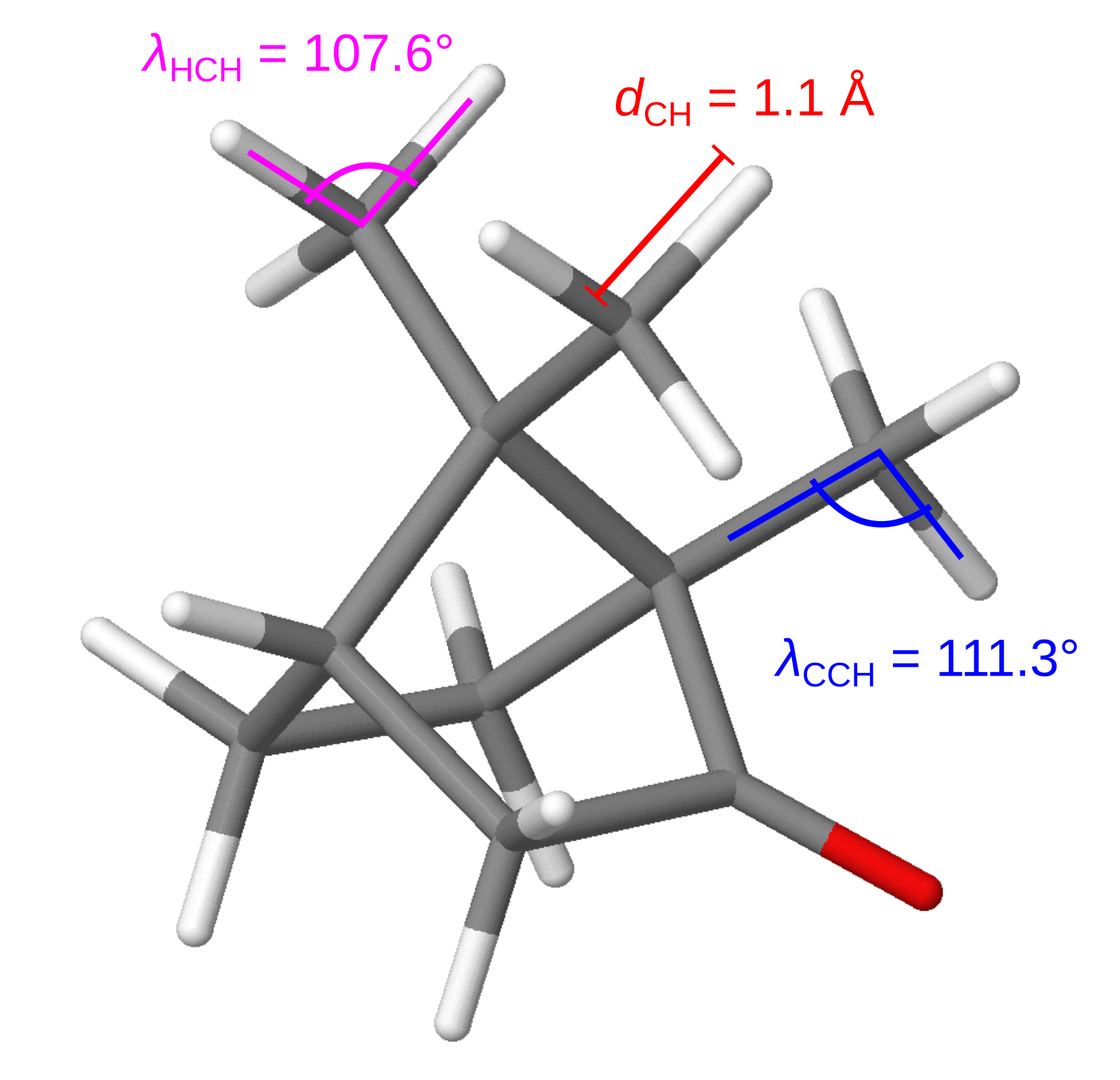}
\caption{\label{fig_sm_ideal}Ideal camphor geometry with a 120-degree periodic rotation of the methyl groups. The C-H bonds in the three methyl groups are normalized to their average relaxed bond length of $d_\mathrm{CH} = 1.1$ \AA, bond angle $\lambda_\mathrm{CCH} = 111.3^\circ$ and the angle between H atoms $\lambda_\mathrm{HCH} = 107.6^\circ$.}
\end{figure}

\clearpage

\section{Convergence of the 6D surrogate model}

We evaluate the convergence of the surrogate model of the 6D PES with respect to the coordinates and adsoprtion energy of the identified local minima. For a reliable identification of all the minima, we converge the complete PES, not only the global minimum. In this process, we acquire new energy points with BOSS and DFT in batches. We follow the convergence of the 6D model by identifying the minima in the model after each batch. Before the local minima search, the acquired energy points are duplicated according to the translational symmetry in the orthogonal unit cell of the Cu(111) surface.

The minima are identified using the built-in local minima search functionality of BOSS. From each acquisition point, BOSS starts a minimizer, which traverses the landscape following the gradient to locate an energy minimum. The minimizers apply the limited-memory Broyden–Fletcher–Goldfarb–Shanno (L-BFGS) optimization algorithm. With this method, multiple minimizers typically end up in the same minimum. The duplicate minima are not removed, since they provide information about the surrogate model, that is, how large region of the phase space a particular minimum occupies. Due to the varying number of acquisition points in different models, the number of employed minimizers varies between models. We analyzed the convergence of the 6D PES with 3 different models, constructed with 1218, 1380, and 1420 energy points (referred to as M1218, M1380, and M1420). In the local minima search, we consider the reliably identified minima to be the points, into which several minimizers have ended up after traversing the landscape. 

\begin{figure*}[h]
\includegraphics{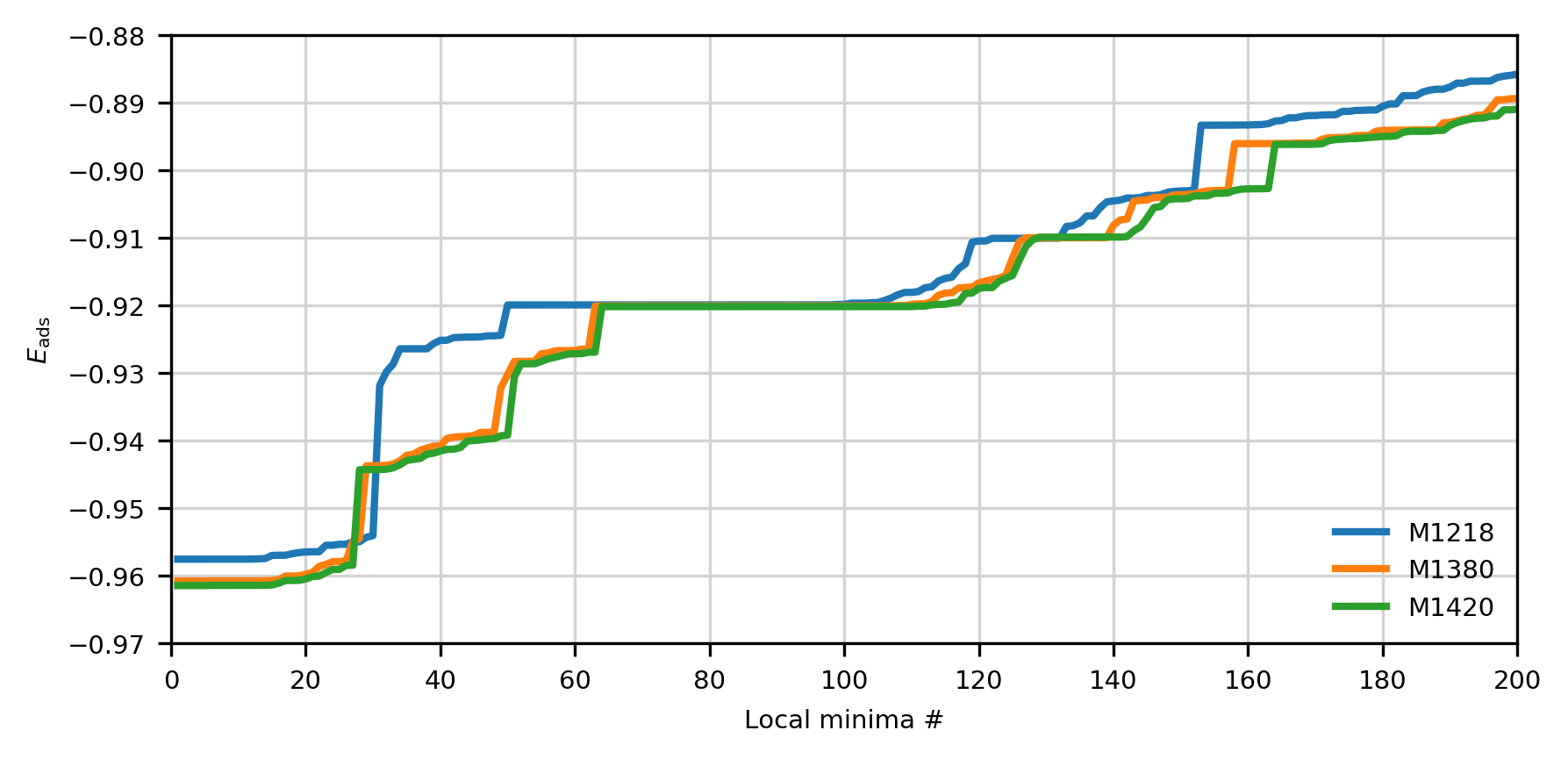}
\caption{\label{fig_sm_conv}Adsorption energy ($E_\textrm{ads}$) of 200 lowest minima in 3 different surrogate models of the 6D PES, constructed with 1218, 1380, and 1420 energy points. The local minima are identified as points of identical energy, found by multiple minimizers, which are shown as energy plateaus in the graph.}
\end{figure*}

We identify the minima by investigating the adsorption energy of each minimizer, sorted by energy (Fig.~\ref{fig_sm_conv}). In this graph, the minima are shown as energy plateaus of varying lengths. The longer the plateau, the more minimizers have ended up in the particular minimum. The varying number of minimizers in each model shows as a horizontal shift of the graph between the 3 models. From this analysis, we conclude that the converged model is M1380. We then proceed by extracting the structures in the identified minima, verify them with full relaxation with DFT, and perform further analysis.

\clearpage

\section{Coordinates of camphor in the predicted and relaxed stable structures}

We confirm the accuracy of the predicted stable structures, identified with Bayesian Optimization Structure Search (BOSS) in the minima of the 6D potential energy surface (PES), with a full relaxation with density-functional theory (DFT). Here, we compare the 3 translational $(x, y, z)$ and 3 rotational $(\alpha, \beta, \gamma)$ coordinates of camphor in the minima of the 6D PES (Tab.~\ref{tab_sm_pred_coords}) to the corresponding coordinates after full relaxation of the structures (Tab.~\ref{tab_sm_relax_coords}). We then evaluate the structural changes in the relaxation via difference of the coordinates before and after the relaxation (Tab.~\ref{tab_sm_coorddiff}).

\begin{table*}[h]
\caption{\label{tab_sm_pred_coords}Coordinates of camphor in the identified stable structures (i.e. structures with building block approximation), which correspond to the minima of the 6D PES.}
\setlength{\tabcolsep}{10pt}
\begin{tabular}{ccc|ccc|ccc}
 \hline
& $x$ (frac.) & $y$ (frac.) & $x$ (\AA) & $y$ (\AA) & $z$ (\AA) & $\alpha$ (deg) & $\beta$ (deg) & $\gamma$ (deg) \\
\hline
Ox1 & $-0.271$ & $-0.301$ & $-0.695$ & $-1.337$ & $4.174$ & $12.7$ & $5.8$ & $-153.6$ \\
Ox2 & $0.206$ & $0.326$ & $0.528$ & $1.451$ & $4.454$ & $-85.7$ & $161.6$ & $2.2$ \\
Ox3 & $-0.393$ & $-0.018$ & $-1.009$ & $-0.078$ & $4.404$ & $4.8$ & $25.5$ & $173.7$ \\
Ox4 & $0.183$ & $0.326$ & $0.471$ & $1.451$ & $4.603$ & $31.5$ & $3.8$ & $-102.4$ \\
Ox5 & $0.004$ & $-0.005$ & $0.011$ & $-0.023$ & $4.973$ & $58.6$ & $46.8$ & $-58.4$ \\
\hline
Hy1 & $-0.001$ & $0.000$ & $-0.003$ & $0.000$ & $5.000$ & $-111.1$ & $32.0$ & $-175.0$ \\
Hy2 & $-0.463$ & $-0.430$ & $-1.190$ & $-1.912$ & $4.071$ & $-98.4$ & $-90.9$ & $-81.3$ \\
Hy3 & $0.000$ & $0.000$ & $-0.001$ & $0.000$ & $5.023$ & $178.8$ & $33.8$ & $-34.9$ \\
\hline
\end{tabular}
\end{table*}

\begin{table*}[h]
\caption{\label{tab_sm_relax_coords}Coordinates of camphor in the stable structures after full relaxation with DFT. The rotational coordinates of structure Hy2 are not uniquely defined (denoted with~***) and therefore omitted here. A visual comparison of structure Hy2 before and after the relaxation is shown in Fig.~\ref{fig_sm_Hy2}.}
\setlength{\tabcolsep}{10pt}
\begin{tabular}{ccc|ccc|ccc}
 \hline
& $x$ (frac.) & $y$ (frac.) & $x$ (\AA) & $y$ (\AA) & $z$ (\AA) & $\alpha$ (deg) & $\beta$ (deg) & $\gamma$ (deg) \\
\hline
Ox1 & $-0.281$ & $-0.306$ & $-0.721$ & $-1.362$ & $4.127$ & $12.9$ & $4.2$ & $-152.3$ \\
Ox2 & $0.198$ & $0.374$ & $0.508$ & $1.662$ & $4.379$ & $-93.5$ & $169.0$ & $-0.4$ \\
Ox3 & $-0.411$ & $-0.020$ & $-1.055$ & $-0.088$ & $4.130$ & $9.1$ & $15.6$ & $178.3$ \\
Ox4 & $0.292$ & $0.296$ & $0.749$ & $1.318$ & $4.424$ & $38.8$ & $-4.3$ & $-96.6$ \\
Ox5 & $0.018$ & $-0.017$ & $0.045$ & $-0.077$ & $4.778$ & $56.8$ & $44.8$ & $-58.6$ \\
\hline
Hy1 & $0.221$ & $0.023$ & $0.567$ & $0.101$ & $4.339$ & $-89.8$ & $20.9$ & $-171.9$ \\
Hy2 & $-0.490$ & $-0.474$ & $-1.257$ & $-2.109$ & $3.964$ & $***$ & $***$ & $***$ \\
Hy3 & $-0.001$ & $-0.001$ & $-0.002$ & $-0.005$ & $5.030$ & $178.8$ & $34.0$ & $-34.8$ \\
\hline
\end{tabular}
\end{table*}

\begin{table*}[h]
\caption{\label{tab_sm_coorddiff}Structural changes in the identified stable adsorbates, comparing the predicted structures (i.e. structures with building block approximation) to the structures after full relaxation with DFT. Change in the location and orientation of camphor is given with respect to translational and rotational coordinates, ($\Delta x, \Delta y, \Delta z$) and ($\Delta \alpha, \Delta \beta, \Delta \gamma$), respectively. Change in the internal geometry of camphor is given as root-mean-square deviation of the atomic positions ($\delta^\textrm{A}$) and the mean deviation of bond lengths ($\delta^\textrm{B}$). Changes in the rotational coordinates of structure Hy2 are not uniquely defined (denoted with~***) and therefore omitted here. A visual comparison of structure Hy2 before and after the relaxation is shown in Fig.~\ref{fig_sm_Hy2}.}
\setlength{\tabcolsep}{9pt}
\begin{tabular}{cccc|ccc|cc}
 \hline
& $\Delta x$~(\AA) & $\Delta y$~(\AA) & $\Delta z$~(\AA) & $\Delta \alpha$~(deg) & $\Delta \beta$~(deg) & $\Delta \gamma$~(deg) & $\delta^\textrm{A}$~(\AA) &  $\delta^\textrm{B}$~(\AA) \\
\hline
Ox1 & $-0.026$ & $-0.025$ & $-0.047$ & $+0.2$ & $-1.6$ & $+1.3$	& 0.033	& 0.0036 \\
Ox2 & $-0.020$ & $+0.211$ & $-0.074$ & $-7.8$ & $+7.4$ & $-2.6$	& 0.136	& 0.0040 \\
Ox3 & $-0.046$ & $-0.009$ & $-0.274$ & $+4.3$ & $-9.9$ & $+4.6$	& 0.142	& 0.0036 \\
Ox4 & $+0.278$ & $-0.133$ & $-0.178$ & $+7.3$ & $-8.1$ & $+5.8$	& 0.180	& 0.0041 \\
Ox5 & $+0.035$ & $-0.055$ & $-0.195$ & $-1.8$ & $-2.0$ & $-0.2$	& 0.072	& 0.0036 \\
\hline
Hy1 & $+0.570$ & $+0.101$ & $-0.661$ & $+21.3$ & $-11.1$ & $+3.1$ & 0.353 & 0.0020 \\
Hy2 & $-0.068$ & $-0.198$ & $-0.108$ & $***$ & $***$ & $***$	& 0.136 & 0.0025 \\
Hy3 & $-0.001$ & $-0.005$ & $+0.007$ & $-0.0$ & $+0.2$ & $+0.1$ & 0.010 & 0.0015 \\
\hline
\end{tabular}
\end{table*}

We applied BOSS to solve the rotational coordinates of the relaxed structures. With a 3D search, BOSS acquired different molecular orientations and identified the rotation $R(\alpha, \beta, \gamma)$ that produces a matching orientation with the relaxed structure.

\begin{figure}[h]
\includegraphics[width=12cm]{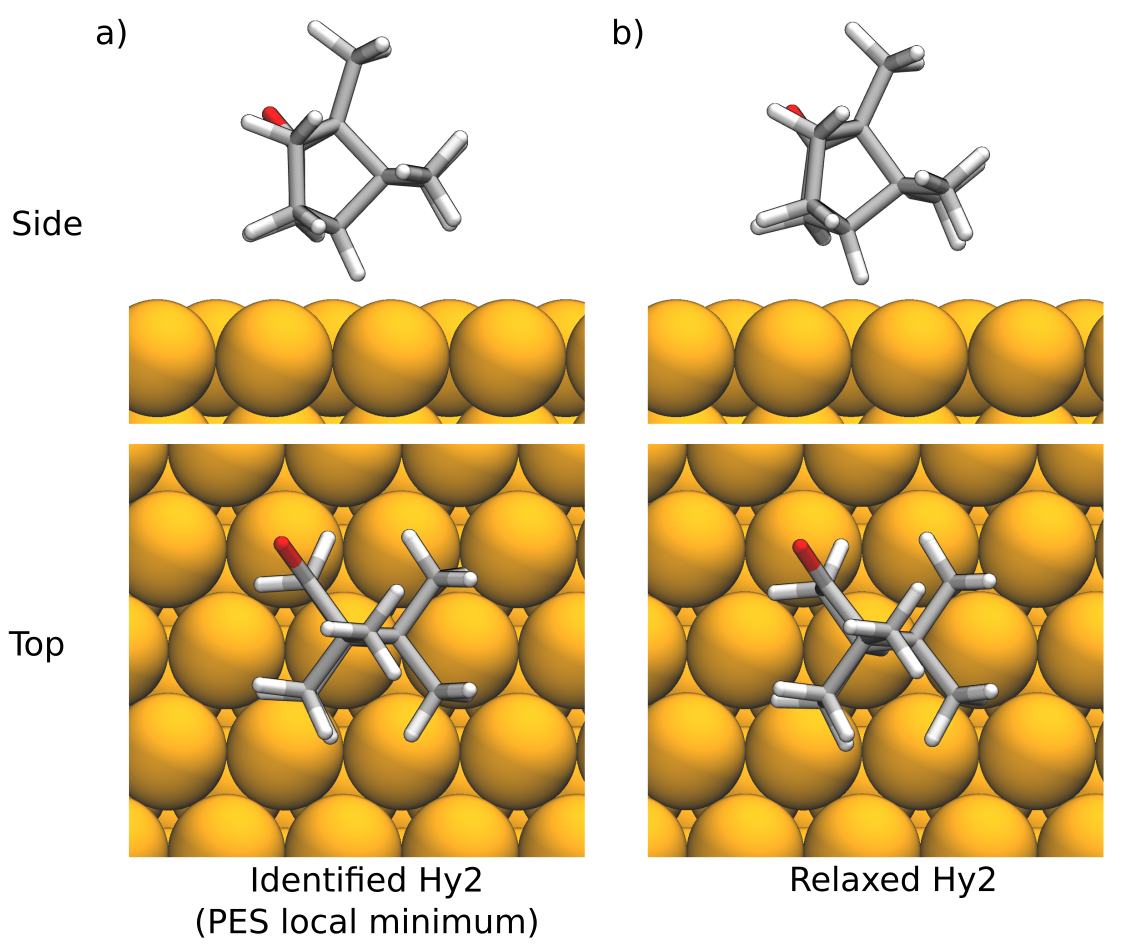}
\caption{\label{fig_sm_Hy2}a) Stable structure Hy2, identified in the local minimum of the 6D PES (i.e. with building block approximation), and b) after full relaxation with DFT.}
\end{figure}

With structure Hy2, the rotational coordinates are not uniquely defined. Due to the $\beta$ rotation angle of ca. $-90^\circ$ in Hy2, the $\alpha$ and $\gamma$ rotations are coupled, such that identical orientations can be produced with various different values of $\alpha$ and $\gamma$. We have therefore omitted the rotational coordinates of the relaxed structure Hy2 in Tab.~\ref{tab_sm_relax_coords} and Tab.~\ref{tab_sm_coorddiff} (denoted with~***). Nevertheless, we verified structure Hy2 with visual comparison before and after the relaxation (Fig.~\ref{fig_sm_Hy2}), which clearly shows that the structural changes in the relaxation are minimal.

\end{document}